\documentclass[aps,twocolumn,pra,superscriptaddress,showpacs,tightenlines]{revtex4}
\usepackage{amssymb}
\usepackage{amsmath}
\usepackage{graphicx}
\usepackage{epsfig}
\usepackage{txfonts}
\usepackage{subfigure}
\usepackage{amsfonts}
\usepackage{CJK}
\begin{document}

\begin{CJK*}{GBK}{song}
\title{Quantum-discord-induced superradiance and subradiance in a system of two separated atoms}

\author{Shi-Qing Tang}
\affiliation{Key Laboratory
of Low-Dimensional Quantum Structures and Quantum Control of
Ministry of Education,  and Department of Physics, Hunan Normal
University, Changsha 410081, China}
\author{Ji-Bing Yuan}
\affiliation{Key Laboratory of Low-Dimensional Quantum Structures
and Quantum Control of Ministry of Education,  and Department of
Physics, Hunan Normal University, Changsha 410081, China}
\author{Xin-Wen Wang}
\affiliation{Key Laboratory of Low-Dimensional Quantum Structures
and Quantum Control of Ministry of Education,  and Department of
Physics, Hunan Normal University, Changsha 410081, China}
\affiliation{Department of Physics and Electronic Information
Science, Hengyang Normal University, Hengyang 421002, China}
\author{Le-Man Kuang\footnote{Author to whom any correspondence should be
addressed. Email: lmkuang@hunnu.edu.cn}}
\affiliation{Key Laboratory
of Low-Dimensional Quantum Structures and Quantum Control of
Ministry of Education,  and Department of Physics, Hunan Normal
University, Changsha 410081, China}


\begin{abstract}
We investigate collective radiant properties of two separated atoms
in  $X$-type quantum states. We show that quantum correlations
measured by quantum discord (QD) can induce and enhance
superradiance and subradiance in the two-atom system even though in
the absence of interatomic quantum entanglement. We also explore
quantum statistical properties of photons in the superradiance and
subradiance by addressing the second-order correlation function. In
particular, when the initial state of the two separated atoms is the
Werner state with non-zero QD, we find that radiation photons in the
superradiant region exhibit the nonclassical sub-Poissonian
statistics and the degree of the sub-Poissonian statistics increases
with increasing of the QD amount, while radiation photons in the
subradiant region have either the sub-Poissonian or super-Poissonian
statistics depending on the amount of QD and the directional angle.
In the subradiant regime we predict the QD-induced photon statistics
transition from the super-Poissonian to sub-Poissonian statistics.
These results shed a new light on applications of QD as a quantum
resource.

\end{abstract}


\maketitle \narrowtext
\end{CJK*}

\section{\label{Sec:1}Introduction}

Quantum discord (QD) \cite{oll,hen} is a measure of nonclassical
correlations which include not only quantum correlations with
entanglement but also entanglement-free quantum correlations that
may occur in separable states. QD captures the nonclassical
correlations, more general than entanglement, that can exist between
parts of a quantum system even if the corresponding quantum
entanglement does vanish. QD  is considered to be a more general
resource of quantum advantage than quantum entanglement in quantum
information
processing~\cite{Datta2008,mer,Lanyon2008,Roa2011,Li2012,Datta2009,Madhok2012,dak,pia}.
QD has been investigated in a wider context
\cite{pat,mad,cav,str,chu,dil,wer,wan,maz,li,yua1,yua2,xu}
including
 entanglement-free quantum computation \cite{Datta2008}, quantum
communication \cite{dak,mad,cav,str,chu}, and quantum phase
transitions in many-body physics \cite{dil,wer,wan,maz,li,yua2}. In
this paper, we add superradiance and subradiance to the list of QD
applications. We show that QD can induce superradiance and
subradiance in a system of two separated atoms.

Superradiance, first introduced by Dicke \cite{93PR99}, is a
coherent spontaneous emission generated by the cooperative effect of
many atoms, molecules, or nuclei. At this point, all the atoms
simultaneously interact with a common radiation field and exhibit a
collective behavior due to the coherent superposition. Ordinary
spontaneous emission is usually incoherent spontaneous radiation,
because the phases of radiations of different atoms have no any
correlation with each other. As a consequence, the emitted light
pulse is a simple algebraic addition of the independent radiation
intensity, and its shape  decays exponentially. The shape of
superradiant pulse, however, does not take on exponential decay and
has a peak after a certain delay. Note that superradiant emission in
certain directions is accompanied by subradiant emission in other
different directions \cite{93PR99,45CJP1833,strou,54PRL1917}. In the
original Dicke model, the superradiance was shown to be a
consequence of the macroscopic dipole moment produced by the
collective coherent behavior of numerous atoms in a local system. It
has been shown that the larger the macroscopic dipole moment is, the
more visible the superradiance phenomenon will be. Since the
superradiance was observed firstly by Skribanowitz \emph{et al}
\cite{30PRL309}, it has attracted considerable interest and been
extensively investigated \cite{76PRL2049,96PRL010501,
17LaserPhys635, 325Science99, 102PRL143601, 81PRA053821, 57JMO1311,
85PRA013821,77PRA033844, 82PRA023827, 108PRL123602,328Science1248,
482Nature199}, due to its wide applications such as generating
$X$-ray lasers with higher power. Much of the aforementioned
literatures were focused on the ensembles of atoms with inherent
complexities associated with ensembles, where the interatomic
distance was considered to be shorter than the emission wavelength.
Recently, Wiegner, Zanthier, and Agarwal \cite{84PRA023805}
investigated the superradiance phenomenon of an $N$-atom system
prepared in the canonical $W$ state \cite{62PRA062314}. It was shown
that superradiance occurs even if the interatomic distance is larger
than the wavelength of the emitted photons, due to the existence of
quantum entanglement among these atoms. It should be pointed out
that such a system has no dipole moment due to the fact that the
distance between any two atoms is larger than the emission
wavelength and the dipole-dipole interactions are negligible.

In this paper, we study the radiative characteristics of emission
for a two-atom system with quantum correlations characterized by the
QD. We show that two atoms in a nonzero-QD state can create
superradiance and subradiance even without interatomic entanglement,
that is, entanglement-free QD can induce superradiance and
subradiance. The rest of this paper is organized as follows. In
Sec.~\ref{Sec:2}, we investigate the radiative characteristics of
the emission from the two two-level atoms initially in a $X$-type
state, and demonstrate the existence of the QD-induced super- and
sub-radiance phenomena. In Sec.~\ref{Sec:3}, we analyze quantum
statistical properties of emitted photons and indicate that photons
in super- and sub-radiance may display different statistical
statistics. In particular, it is found that the QD effect can induce
nonclassical photon statistics and the transition from the
super-Poissonian to sub-Poissonian statistics.  Finally, this work
is concluded in Sec.~\ref{Sec:4}.

\section{\label{Sec:2} Superradiance and subradiance of two separated atoms in a $X$-type state}

In this section, we investigate radiation properties of a quantum
system consisting of two separated atoms. We show that quantum
discord can induce superradiance and subradiance of the two-atom
system which is initially in a $X$-type state. The system under our
consideration is shown in Fig.~1. It consists of two identical
two-level atoms which are localized at positions $z_1$ and $z_2$,
respectively, with $|z_1-z_2|=l$. We assume that  $kl >1$ with the
wave number $k=2\pi/\lambda$, so that the dipole-dipole interaction
can be neglected.

\begin{figure}[tbp]
\includegraphics[width=8.0cm,height=5cm]{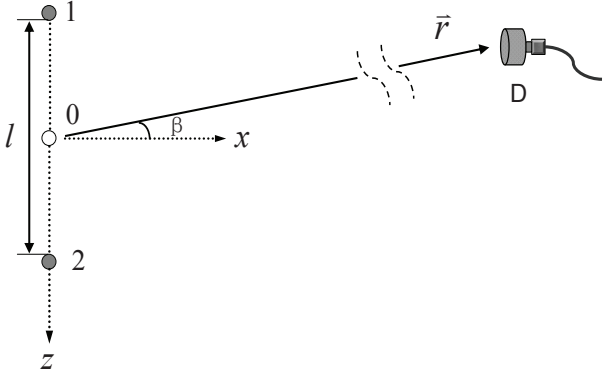}
\caption{Schematic diagram of observing superradiance and
subradiance of two separated atoms. Two identical two-level atoms
are arranged along the ${z}$-axis with coordinates $z_1=-l/2$ and
$z_2=l/2$, respectively. A detector is placed at position $\vec{r}$
in the far field of the source with a directional angle $\beta$.}
\label{fig1}
\end{figure}

We now consider a detector placed at position ${\vec r}$ in the
far-field region of the two atoms to measure the emission intensity
\begin{equation}
 \I =\langle\hat{E}^{-}\, \hat{E}^{+} \rangle,\label{a}
\end{equation}
where $\hat{E}^{+}$ and $\hat{E}^{-}$ [$=(\hat{E}^{+})^\dagger$]
are, respectively, the positive and negative frequency parts of the
electric field operator. $\hat{E}^{+}$ is given by \cite{Agarwal}
\begin{equation}
\hat{E}^{+} \sim \frac{e^{ikr}}{r} \sum_{j=1}^{2} \vec{n}
\times(\vec{n} \times \vec{\mu}) e^{-i\alpha_{j}}\hat{\sigma}^{-}_j\
 , \label{b}
\end{equation}
where $\vec{n} = \frac{\vec{r}}{r}$ is the unit vector, $\vec{\mu}$
is the dipole moment of the transition
$|e\rangle\leftrightarrow|g\rangle$, $\hat{\sigma}^{-}_j =
|g\rangle_j\langle e|$ denotes the dipole operator, and $\alpha_j= k
\vec{n}\cdot\vec{z}_{j} = jkl\sin\beta$ denotes the relative optical
phase accumulated by a photon emitted at $\vec{z}_j$ and detected at
${\vec r}$. For clarity, we assume $\vec{\mu}$ is along the $y$-axis
direction and $\vec{n}$ is in the $x-z$ plane, which means
$\vec{\mu}\cdot \vec{n} = 0$. Under these conditions and omitting
all constant factors, Eq.~(\ref{b}) reduces to \cite{84PRA023805}
\begin{equation}
\hat{E}^{+} \sim \sum_{j=1}^{2} e^{-i\alpha_{j}}
\hat{\sigma}^{-}_j\,. \label{c}
\end{equation}
The field is now dimensionless and hence all intensities would be
dimensionless. The actual values can be obtained by multiplying the
emission produced by a single excited atom.

The initial state of the two atoms is taken to be a $X$-type state
given by \cite{yu,luo,ali}
\begin{eqnarray}
\hat{\rho}&=&\frac{1}{4}\left(\hat{\mathcal{I}}_{12}+\underset{i=x,y,z}{\sum
}c_{i}\hat{\sigma}_{i}^{1}\otimes \hat{\sigma} _{i}^{2}\right)
\nonumber \\
&=&\frac{1}{4}\left(
\begin{array}{cccc}
1+c_{z} & 0 & 0 & c_x-c_y \\
0 & 1-c_{z} & c_x+c_y & 0 \\
0 & c_x+c_y & 1-c_{z} & 0 \\
c_x-c_y & 0 & 0 & 1+c_{z}
\end{array}
\right) , \label{5}
\end{eqnarray}
where $\hat{\mathcal{I}}_{12}$ is the identity operator in the
Hilbert space of two qubits, $\hat{\sigma} _{i}^{1}$ and
$\hat{\sigma} _{i}^{2}$ are, respectively, the Pauli matrices of
qubits $1$ and $2$, and the real numbers $c_{i}$ ($0\leqslant
\left\vert c_{i}\right\vert \leqslant 1$) satisfy the conditions
that the state-density matrix $\hat{\rho}$ is positive and the trace
of it is unit. Here we have used the conventional denotations
$|e\rangle=|1\rangle=[1,0]^T$ and $|g\rangle=|0\rangle=[0,1]^T$.
Such a state has the maximally mixed marginal, that is, $\hat{\rho}
_{1(2)}=\hat{\mathcal{I}}_{1(2)}/2$.

According to Eqs.~(\ref{a}) and (\ref{c}), we can obtain the
dimensionless radiation intensity of the two atoms in the $X$-type
state of Eq.~(\ref{5}),
\begin{equation}
I=\mathrm{tr}(\hat{\rho} \hat{E}^{-}\hat{E}^{+})=1+\frac{1}{2}\left(
c_{x}+c_{y}\right) \cos \left( kl\sin \beta \right) .\label{d}
\end{equation}
It can be seen that the radiation intensity $I$ can be larger or
less than unit. Note that when the atoms emit photons at random, the
radiation intensity is one, due to the fact that the probability of
the two atoms occupying the excited state is
\begin{equation}
 \mathrm{tr}\left[ (|e\rangle_{1}\langle e| +|e\rangle_{2}\langle e|)\hat{\rho} \right]=1.\label{7}
\end{equation}
Thus, Eq. (\ref{d}) indicates that it is possible to create
superradiance ($I>1$) and subradiance ($I<1$) for the two-atom
system in the $X$-type state.

In the following, we give a detailed discussion on the case of
$c_x=c_y=c_z=-c$. In this case, the $X$-type state in Eq.~(\ref{5})
reduces to a Werner state \cite{wern}, characterized by a single
parameter, given by
\begin{equation}
 \hat{\rho} =\frac{1}{4}\left(
\begin{array}{cccc}
1-c & 0 & 0 & 0 \\
0 & 1+c & -2c & 0 \\
0 & -2c & 1+c & 0 \\
0 & 0 & 0 & 1-c%
\end{array}%
\right) ,\label{8}
\end{equation}
where $ c\in[0,1]$. The nonclassical correlation of the Werner state
can be quantified by quantum discord \cite{oll,hen}, which is
calculated to be \cite{luo,ali}
\begin{eqnarray}
 D\left( c\right)
&=&\frac{1-c}{4}\log_{2}(1-c)-\frac{1+c}{2}\log_{2}(1+c)\nonumber\\
&&+\frac{1+3c}{4}\log_{2}(1+3c) .\label{9}
\end{eqnarray}
The amount of entanglement of the Werner state can be measured by
concurrence \cite{80PRL2245}, which is given by
\begin{equation}
C(c)=\texttt{max} \left\{0, \frac{1}{2}(3c-1)\right\}.
\end{equation}
Obviously, the concurrence is zero for $0\leqslant c\leqslant 1/3$.
This indicates that the Werner state in the parameter regime of
$0\leqslant c\leqslant 1/3$ is a separable quantum state without
entanglement. However, from Eq. (9) we can find that the QD of the
Werner state in the parameter regime of $0\leqslant c\leqslant 1/3$
is nonzero. Hence, the Werner state with $0\leqslant c\leqslant 1/3$
is such a quantum state with quantum correlation but without quantum
entanglement. In particular, at the entanglement-transition point of
$c=1/3$, we have the QD with $D\equiv D_c \approx 0.126$. The
numerical analysis indicates that both the QD amount $D$ and the
concurrence $C$ increase with the increasing of the state parameter
$c$ in the range $c> 1/3$ \cite{luo,ali}.

\begin{figure}[tbp]
\includegraphics[width=8cm,height=7cm]{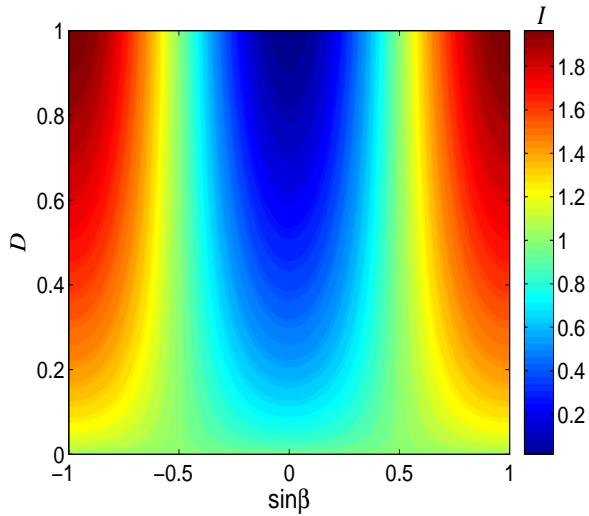}
\caption{(Color online) Plot of the radiation intensity ${I}$ with
respect to the QD amount ${D}$ and the direction angle parameter
$\sin \beta$, where $kl=\pi$.} \label{fig2}
\end{figure}

For the Werner state in Eq.~(\ref{8}), the radiation intensity given
by Eq. (\ref{d}) becomes
\begin{equation}
 I=1-c\cos \left(kl\sin\beta\right) ,\label{e}
\end{equation}
from which we can see that the radiation intensity depends on not
only the interatomic distance $l$ and the detection angle $\beta$
but also the state parameter $c$ for a given wave vector $k$. We
have plotted the radiation intensity with respect to the QD and the
direction angle parameter $\sin\beta$ in Fig. 2 which indicates the
existence of the QD-induced superradiance and subradiance in
different angle regimes. From Fig. 2 we can see that the nearby
region along the $\pm x$-axis direction with $|\sin \beta |< 0.5$ is
the subradiance region with $I< 1$ while the nearby region along the
$\pm z$-axis direction with $|\sin \beta|> 0.5$ is the superradiance
region with $I>1$.

In order to clearly see the QD influence on the radiation intensity,
in Fig. 3 we have plotted the radiation intensity with respect to
the QD for the directional parameter $\sin \beta$ taking $0$ and
$1$, respectively. From Fig. 3 we can clearly see the QD-induced
superradiant and subradiant phenomena. The superradiance appears in
the branch of $\sin \beta =1$ while the subradiance occurs in the
branch of $\sin \beta= 0$. The superradiance (subradiance) is
enhanced with the increase of the QD of the Werner state since the
radiation intensity $I$ monotonically increases (decreases) with
increasing of the QD value $D$. This shows that the QD can enhance
the superradiance and subradiance.

\begin{figure}[tbp]
\includegraphics[width=8cm,height=7cm]{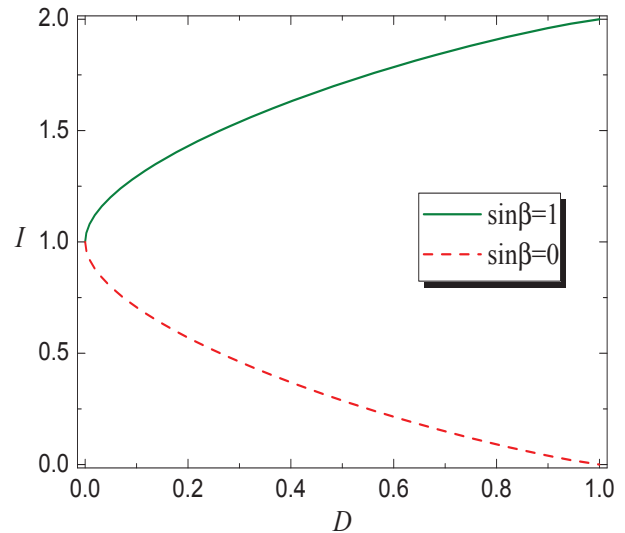}
\caption{(Color online) The radiated intensity ${I}$ versus the QD
amount ${D}$ with $\sin\beta$ taking $1$ (the solid  line) and $0$
(the dashed line), respectively. Here we take $kl=\pi$.}
\label{fig3}
\end{figure}

A notable consequence could be obtained from Fig. 3 that quantum
correlation is at least a necessary condition for generating
superradiant and subradiant emission for the two-separated-atom
system under our consideration. Note that when $0<c\leqslant 1/3$,
the Werner state has no entanglement while its quantum discord is
larger than zero ($0<D\leqslant D_c$) \cite{luo,ali}. This implies
that entanglement is not a necessary condition for yielding
superradiance and subradiance, in contrast to the case where the
atoms are initially in a pure entangled state, supporting the
necessity of entanglement \cite{84PRA023805} to produce
superradiance and subradiance. Therefore, we can conclude that
quantum correlations without entanglement, i.e., the
entanglement-free QD, can induce and enhance the superradiance and
subradiance of two separated atoms.

\section{\label{Sec:3}Quantum statistical properties of radiated photons}

This section aims at revealing the quantum statistical behavior of
radiated photons by analyzing the second-order correlation function
of the radiation intensity. We only pay attention to the photon
statistics in superradiance and subradiance. It will be shown that
quantum statistical properties of photons in the superradiance are
different from those in the subradiance. More interestingly, we find
that the QD can induce nonclassical photon statistics \cite{dav} and
the statistical property could jump from the classical
super-Poissonian to the nonclassical sub-Poissonian in some angle
regimes.

The second-order correlation function of the radiation intensity is
defined as
\begin{equation}
 g^{\left(2\right)}(0)=\frac{\mathrm{tr}\left( \hat{\rho} \hat{E}^{-^{2}}\hat{E}^{+^{2}}\right)}{%
\left[ \mathrm{tr}\left( \hat{\rho}
\hat{E}^{-}\hat{E}^{+}\right)\right] ^{2}} ,\label{11}
\end{equation}
where $\rho$ is the density operator of the radiation field. With
the second-order correlation function, one can distinguish between
the possibly different statistical regimes. In particular,
$g^{(2)}(0) <1$ denotes the sub-Poissionian statistics of the
radiation field, $g^{(2)}(0)
>1$ means the sub-Poissionian statistics, and $g^{(2)}(0) =1$
reflects the Poissionian statistics.

After a straightforward calculation we obtain the second-order
correlation function of the radiation intensity of the two atoms
being initially in the Werner state of Eq.~(\ref{8}),
\begin{equation}
g^{\left( 2\right) }(0)=\frac{1-c}{%
\left[ 1-c\cos (kl\sin \beta) \right]^{2}}. \label{12}
\end{equation}
When $c=0$, we get $g^{(2)}(0)= 1$ meaning that the radiation of the
system degenerates to spontaneous emissions of the two independent
atoms. This point can be understood from the fact that QD vanishes
in this case, that is, there is no any quantum correlation between
the two atoms.

\begin{figure}[tbp]
\includegraphics[width=8cm,height=7cm]{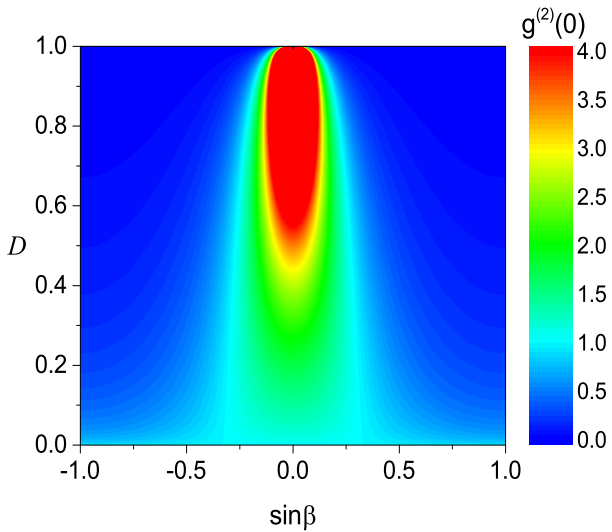}
\caption{(Color online) The second-order correlation function of
radiation photon with respect to the QD amount $D$ and the
directional angle $\beta$  when the $kl=\pi$.} \label{fig4}
\end{figure}

In order to observe how the quantum correlations affect quantum
statistical properties of the radiation photons in the superradiant
and subradiant regimes, we perform numerical simulations for the
influence of the QD of the atomic initial state on the second-order
correlation function of the radiation intensity. In Fig. 4, we have
plotted the second-order correlation function with respect to the
amount of QD $D$ and the directional angle parameter $\sin \beta$.
Combining Fig. 4 with Fig. 2, we can see that photons in the
superradiant regime of $|\sin\beta| > 0.5$ exhibit the
sub-Poissonian statistics with $g^{\left( 2\right)}(0) <1$. In the
subradiant regime of $|\sin\beta| < 0.5$, the quantum statistical
properties of the radiated photons become more complicated as shown
below. The second-order correlation function may have an inflexion
and the photon statistics may change sharply with the variation of
the QD amount. These effects lead to an interesting phenomenon that
QD can induce the photon statistics transition from the classical
super-Poissonian to the sub-Poissonian statistics. Such a phenomenon
can be observed more clearly in Fig.~5 which gives the second-order
correlation function as a function of the QD amount $D$ with $\sin
\beta=0.2$. It can be seen that the transition point happens at the
position of $D=D_t\approx0.87$. Photon statistics in the range
$0<D<D_t$ is the super-Poissonian while photon statistics in the
range $D_t<D<1$ is the sub-Poissonian. At the transition point
$D_t$, photon statistics is the Poissonian. Therefore, we can
conclude that quantum statistical behaviors of radiation photons can
be manipulated by changing the interatomic quantum correlations for
some fixed direction angles, and that the QD can induce the photon
statistics transition from classical super-Poissonian to the
sub-Poissonian statistics.

\begin{figure}[tbp]
\includegraphics[width=8cm,height=7cm]{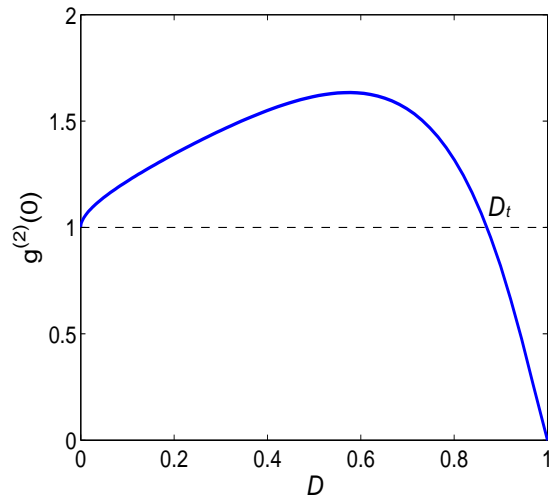}
\caption{(Color online) The second-order correlation function of
radiation photon with respect to the QD amount $D$  when $\sin
\beta=0.20$ and $kl=\pi$.} \label{fig5}
\end{figure}

\section{\label{Sec:4} Concluding remarks}

In conclusion, we have studied collective radiant properties of two
separated atoms in  $X$-type quantum states. It has been shown that
QD can induce and enhance superradiance and subradiance in the
two-atom system. More interestingly, we have found that quantum
correlations without any entanglement can  induce and enhance
superradiance and subradiance. This sheds a new light on
applications of QD as a quantum resource with quantum advantages.
Furthermore, we have investigated quantum statistical properties of
emitted photons in both regions of superradiance and subradiance by
addressing the second-order correlation function. When the initial
state of the two separated atoms is the Werner state with non-zero
QD, it has been found that radiation photons in the superradiant
region exhibit the nonclassical sub-Poissonian statistics and the
degree of the sub-Poissonian statistics increases with increasing of
the QD amount. However, radiation photons in the subradiant region
have different statistical characteristics, they are either the
sub-Poissonian or the super-Poissonian statistics, depending on the
amount of QD and the directional angle. A particularly interesting
finding is that in the subradiance region the QD can induce photon
statistics transition from the super-Poissonian to sub-Poissonian
statistics.

\acknowledgments This work was supported by the 973 Program (Grant
No. 2013CB921804), the NSFC (Grant  No. 11375060, No. 11075050, and
No. 11004050), the PCSIRTU (Grant No. IRT0964), the CPSFFP (Grant
No. 2013T60769), and the HPNSF (Grant No. 11JJ7001).  S. Q. Tang
thanks Dr. Jie-Qiao Liao for useful discussions.

\end{document}